\documentstyle[12pt]{article}
\begin{document} 

\def \o{\overline}
\def \ggff{\gamma\gamma\rightarrow q\bar{q}}
\def \gggg{\gamma\gamma\rightarrow gg}
\def \ggjj{\gamma\gamma\rightarrow jj}

\vspace*{-1in} 
\renewcommand{\thefootnote}{\fnsymbol{footnote}} 
\begin{flushright} 
TIFR/TH/99-51\\
hep-ph/9909567\\
September 1999\\ 
\end{flushright} 
\vskip 65pt 
\begin{center} 
{\Large \bf Large extra dimensions and dijet production \\
\vskip5pt
in $\gamma \gamma$ collisions}\\
\vspace{8mm} 
{\bf 
Dilip~Kumar~Ghosh$^1$\footnote{dghosh@theory.tifr.res.in}, 
Prakash Mathews$^2$\footnote{prakash@imsc.ernet.in}
P.~Poulose$^1$ \footnote{poulose@theory.tifr.res.in}, 
K.~Sridhar$^1$\footnote{sridhar@theory.tifr.res.in}
}\\ 
\vspace{10pt} 
{\sf 1) Department of Theoretical Physics, Tata Institute of 
Fundamental Research,\\  
Homi Bhabha Road, Bombay 400 005, India. \\
\vskip4pt
2) Institute of Mathematical Sciences, CPT Campus, Chennai 600 113, India.} 
 
\vspace{80pt} 
{\bf ABSTRACT} 
\end{center} 
We have studied dijet production in $\gamma \gamma$ collisions with
a view to probing the physics of large extra dimensions. The exchange
of virtual spin-2 Kaluza-Klein excitations is found to modify the
dijet cross-section substantially from its Standard Model value
and allows the effective string scale to be probed to values 
between 2.5 and 6.4 TeV in the unpolarised case. In the case
where the photons are polarised, the limits are seen to improve
by roughly 20\%. Dijet production in $\gamma \gamma$ collisions
is thus shown to be a very effective probe of large extra dimensions.

\vskip12pt 
\noindent 
\setcounter{footnote}{0} 
\renewcommand{\thefootnote}{\arabic{footnote}} 
 
\vfill 
\clearpage 
\setcounter{page}{1} 
\pagestyle{plain}
\noindent There has been tremendous interest recently in the physics of 
large compact extra dimensions \cite{dimo,dimo2,shiu} i.e. a number $n$ of
the dimensions of a higher dimensional string theory are compactified
to a scale $R$ which is very large compared to the Planck scale. The
SM particles are confined to a 3-brane and do not see the effects of these 
dimensions. Only gravitons propagate in the bulk and so the the magnitude of 
$R$ is constrained only by gravitation experiments. The constraints from the
latter experiments are relatively weak \cite{gravexp} and allow the extra 
dimension to be as large as 1 mm. After compactification, an effective
theory of quantum gravity emerges, with the scale of the effective
theory $M_S$ being of the order of a TeV. It is at these low
energies of the order of 1 TeV that we will now expect to see the effects 
of quantum gravity. This very novel idea has far-reaching
consequences~: it is a possible solution to the heirarchy problem
(though the latter manifests itself in a new garb). But, more
interestingly, it is possible to make a viable scenario 
\cite{dimo4} which can survive the existing astrophysical and 
cosmological constraints and predict other interesting consequences like
low-energy unification \cite{dienes, kakushadze}. 
For some early papers on large Kaluza-Klein dimensions, see Ref.~\cite{anto, 
taylor} and for recent investigations on different aspects of the
TeV scale quantum gravity scenario and related ideas, see Ref.~\cite{related}.

The low-energy effective theory that emerges below the scale $M_S$
\cite{sundrum, grw, hlz}, has an infinite tower of massive Kaluza-Klein
states, which contain spin-2, spin-1 and spin-0 excitations. For low-energy 
phenomenology the most important of these are the spin-2 Kaluza-Klein states 
i.e. the infinite tower of massive graviton states which couple to the
SM particles via the energy-momentum tensor, as usual. These couplings
can lead to observable effects at present and future colliders.
There have been several studies exploring these consequences.
Production of gravitions giving rise to characteristic
missing energy or missing $p_T$ signatures 
at $e^+ e^-$ or hadron colliders have been
studied resulting in bounds on $M_S$ which are around 500 GeV to 
1.2 TeV at LEP2 \cite{mpp, keung} and around 600 GeV to 750 GeV at Tevatron 
(for $n$ between 2 and 6) \cite{mpp}. Production of gravitons at
the Large Hadron Collider (LHC) and in high-energy $e^+ e^-$ collisions
at the Next Linear Collider (NLC) have also been considered. 
Virtual effects of graviton exchange in dilepton production at Tevatron 
yields a bound of around 950 GeV \cite{hewett} to 1100 GeV \cite{gmr} on $M_S$, 
in $t \bar t$ production at Tevatron a bound of about 650 GeV is obtained while at the 
LHC this process can be used to explore a range of $M_S$ values upto 
4~TeV \cite{us}. Virtual effects in deep-inelastic scattering at HERA 
put a bound of 550 GeV on $M_S$\cite{us2}, while from jet production at the 
Tevatron strong bounds of about 1.2 TeV  are obtained \cite{us3}.
Pair production of gauge bosons and fermions in $e^+ e^-$ 
collisions at LEP2 and NLC and in $\gamma \gamma$ collisions at the NLC 
\cite{rizzo, agashe, soni, davoudiasl, lee, ks} can probe values
of $M_S$ from 0.5 TeV at LEP2 energies to several TeV at NLC. Virtual
effects of gravitons \cite{davoudiasl2} and real graviton production \cite{gps}
in $e\gamma$ collisions have been shown to probe values of $M_S$ as
high as 5 or 6 TeV. Other processes studied include associated 
production of gravitons with gauge bosons and virtual effects in gauge boson 
pair production at hadron colliders \cite{balazs, cheung, eboli}. 
Higgs production \cite{rizzo2, xhe} and electroweak precision observables 
\cite{precision} in the light of this new physics have also been discussed. 
Astrophysical constraints, like bounds from energy loss for supernovae cores, 
have also been discussed \cite{astro}.

In the present paper we study the effects of virtual graviton exchange
in dijet production in $\gamma \gamma$ collisions at the NLC. The
basic process is the Compton scattering of a low-energy laser beam of
a well-determined frequency off a high energy electron beam \cite{nlc}, 
and the parameters of the photon-photon subprocess are fixed by controlling 
the electron and laser beam parameters. These experiments are planned over 
several steps of $e^+ e^-$ centre-of-mass energy spanning the range between 
500 GeV and 1.5 TeV. Given the relatively clean initial state, very high 
precision is possible in these experiments and, indeed, the degree of 
precision can be enhanced by using polarised initial electrons and laser 
beams. Due to the reach in energy and the high precision, these experiments 
can test the SM very accurately and probe new physics that may lie beyond
the SM. In particular, the NLC can be used very effectively to study
the physics of large extra dimensions and it is with this aim that
we study dijet production in $\gamma \gamma$ collisions at the NLC. 

Since in the $\gamma \gamma$ scattering process, each $\gamma$ is produced
from the electron-laser Compton back-scattering, the energy of the 
back-scattered photon, $E_\gamma$, follows a distribution characteristic 
of the Compton scattering process and can be written in terms of the 
dimensionless ratio $x=E_\gamma/E_e$. It turns out that the maximum value of 
$x$ is about 0.82 so that provides the upper limit on the energy accessible 
in the $\gamma \gamma$ sub-process. To get the full cross-section, the 
$\gamma \gamma$ subprocess cross-section is convoluted with the luminosity 
functions, $f^i_\gamma (x)$, which provide information on the photon flux 
produced in Compton scattering of the electron and laser beams \cite{lumino}. 

In the SM, dijet production in $\gamma \gamma$ collisions takes place
via the usual $t$- and $u$-channel production mechanisms, $\gamma \gamma
\rightarrow q \bar q$, where five massless flavours of quarks are
summed over in the final state. In the presence of large extra dimensions,
the $\gamma \gamma \rightarrow q \bar q$ cross-section gets modified 
because of the $s$-channel exchange of virtual spin-2 Kaluza-Klein 
particles. In the following, we refer to this latter contribution
as the non-SM (NSM) contribution. In addition to this, there is an
entirely new NSM channel that opens up: $\gamma \gamma \rightarrow gg$.
There is no SM contribution in this $gg$ production channel, while in the
$q \bar q$ production channel the SM and the NSM contributions interfere. 
We calculate the cross-section for the polarised case and then obtain the 
unpolarised cross-section by summing over the polarisations of the initial 
photons \footnote{The polarisation of each of the photons is a function of the 
polarisations of the initial electron and laser beams and it is only the 
latter that can be fixed in the experiment. When we present our results for 
the polarised case, we will do so for a fixed choice of these initial 
polarisations}. 

The cross-section for the production of two jets in $\gamma \gamma$
collision can be written in terms of the transverse momentum, $p_T$,
and the rapidity, $y$, of one of the jets as follows:
\begin{equation}
\frac{d^2\sigma}{dp_T^2dy}=\int dx_1\,dx_2\,f(x_1, \xi^2_1)\,f(x_2, \xi^2_2)
    \,s\, \left(\sum_q\frac{d\hat{\sigma}_{(\ggff)}}{dt}+
	\frac{d\hat{\sigma}_{(\gggg)}}{dt}\right)\,\delta (s+t+u) , 
\label{e1}
\end{equation} 
where the $f(x_i, \xi^2_i),\ i=1,2$ are the luminosity functions for each of
the photons where $\xi^2_1$ and $\xi^2_2$ are the Stokes parameters of the 
first and second photons respectively, $d\hat{\sigma}_{(\ggff)}/dt$ and 
$d\hat{\sigma}_{(\gggg)}/dt$ are the subprocess cross-sections for the 
$q \bar q$ and $gg$ channels respectively, and are given in terms of 
helicity amplitudes as follows:
\begin{equation}
\frac{d\hat{\sigma}_{(\ggjj)}}{dt}=\frac{1}{16 \pi s^2}\left(
	\frac{1}{2}(1+\xi^2_1{\xi}^2_2)\,|{\cal M}_j(++)|^2+
	\frac{1}{2}(1-\xi^2_1{\xi}^2_2)\,|{\cal M}_j(+-)|^2 \right) ,
\end{equation}
with the helicity amplitudes are given as: 
\begin{eqnarray}
|{\cal M}_q(++)|^2 &=&0 , \nonumber \\
|{\cal M}_q(+-)|^2 &=&192\pi^2 (s^2+2st+2t^2)\left(-\alpha^2 e_q^4
	\frac{1}{t(s + t)}-\alpha e_q^2\frac{\lambda}{M_S^4}-
	\frac{t(s+t)}{4 M_S^8}\right) , \nonumber \\
|{\cal M}_g(++)|^2 &=&0 , \nonumber \\
|{\cal M}_g(+-)|^2 &=&\frac{64\pi^2}{M_S^8}\left(s^4+4s^3t+6s^2t^2+4st^3+2t^4
\right) .
\label{e2}
\end{eqnarray}
In the above equations, $ {\cal M}_q(\lambda_1, \lambda_2) $ are the helicity 
amplitudes for the subprocess $\gamma \gamma \rightarrow q \bar q$ and 
$ {\cal M}_g(\lambda_1, \lambda_2) $ are those for the subprocess 
$\gamma \gamma \rightarrow gg$, with $\lambda_1$ and $\lambda_2$
denoting the polarisations of the first and the second photons,
respectively. 

\begin{table}
\begin{center}
\begin{tabular}{|c|c|c|c|}
\hline
&\multicolumn{3}{|c|}{ }\\
&\multicolumn{3}{|c|}{$M_S$ in TeV}\\[3mm]
\cline{2-4}
$p_T^{\rm cut}$&&&\\
GeV&$\sqrt{s}_{ee}$=0.5~TeV&1~TeV&1.5~TeV\\
\hline
&&&\\
10&2.579&4.059&5.330\\
35&3.032&4.642&6.007\\
60&3.288&4.992&6.411\\[3mm]
\hline
\end{tabular}
\caption{$M_S$ limits for different values of $e^+e^-$ c.m. energies
in the case of unpolarized beams. }
\end{center}
\end{table}

The non-SM contribution involves two new parameters: the effective string 
scale $M_S$ and $\lambda$ which is the effective coupling at $M_S$. 
$\lambda$ is expected to be of ${\cal O}(1)$, but its sign is not known 
$a\ priori$. In our work we will explore the sensitivity of our results 
to the choice of the sign of $\lambda$. Consequently, the only free
parameter left is $M_S$ so that our results can be directly translated
into quantitative predictions for the reach in $M_S$.

We begin with the results for the unpolarised case. We have computed the 
unpolarised integrated cross-section for three different values of the 
initial $e^+ e^-$ C.M. energy i.e. $\sqrt{s}_{ee}= 500,\ 1000,\ 1500$~GeV
both for the SM and the NSM. In order to obtain the integrated cross-section,
we have integrated the cross-section in Eq.~\ref{e1} over $p_T$ greater
than a chosen $p_T^{\rm cut}$ and over all accessible values of $y$.
From the value of the SM cross-section we derive the $2\sigma$ error band, 
assuming purely statistical errors and assuming an integrated luminosity of 
100 fb${}^{-1}$. Comparing this with the NSM contribution (including the 
interference term in the $q \bar q$ production channel), we obtain the 
$2\sigma$ limits on $M_S$. These limits are displayed in Table 1, as
a function of the $e^+ e^-$ centre-of-mass energy and $p_T^{\rm cut}$.
Stringent limits ranging from about 2.5 TeV at the lowest $\sqrt{s}_{ee}$
to about 6.4 TeV at the highest $\sqrt{s}_{ee}$ are obtained. Also we
find that the bound that we obtain improves as we demand a larger
value of $p_T^{\rm cut}$. Given that the dijet cross-section is
large, for a luminosity of 100 fb${}^{-1}$ we get large event rates
even for a $p_T^{\rm cut}$ of 60~GeV. The value of $p_T^{\rm cut}$ 
may thus be optimised so as to get a larger limit on $M_S$, while
still retaining a large number of dijet events.

\begin{table}
\begin{center}
\begin{tabular}{|c|c|c|c|}
\hline
&\multicolumn{3}{|c|}{ }\\
&\multicolumn{3}{|c|}{$M_S$ in TeV}\\[3mm]
\cline{2-4}
$p_T^{\rm cut}$&&&\\
GeV&$\sqrt{s}_{ee}$=0.5~TeV&1~TeV&1.5~TeV\\
\hline
&&&\\
10&3.029&4.890& 6.530\\
35&3.390&5.334& 6.996\\
60&3.592&5.614& 7.321\\[3mm]
\hline
\end{tabular}
\caption{$M_S$ limits in the polarized case. Oppositely polarized electron 
beams and laser beams are used to obtain the photon beams. Other parameters 
are as stated in the case of Table 1.}
\end{center}
\end{table}

We now discuss the results for the polarised case. As mentioned earlier,
for a given choice of the initial electron and laser polarisations, the photon 
polarisation is fixed once the $x$ value is known. The latter polarisation is 
therefore dependent crucially on the luminosity functions and it is only on the
polarisation of the electron and the laser beams that we have a direct
handle. At the level of the $\gamma \gamma$ subprocess, the helicity 
amplitudes directly reflect the dynamics of the SM and NSM. As can be
seen from Eq.~\ref{e2}, the amplitude where both photons have identical
polarisation ($\lambda_1 = \lambda_2$) vanishes for $both$ the SM and the NSM. 
The helicity structure of the SM and the NSM contributions is similar 
to this extent, the differences between them coming only from the 
non-vanishing helicity amplitudes i.e. when the photons have opposite 
polarisation ($\lambda_1 = - \lambda_2$). For hard photons, the 
sign of $\lambda$ is the same as that of the helicity of the initial 
electron. Therefore, $\lambda_1 = - \lambda_2$ is realised
when the electron beams have opposite helicities i.e. $(\lambda_{e1} = -
\lambda_{e2})$. Further, the luminosity function is such that it peaks
for a certain value of the product $\lambda_e \lambda_l $ (where $\lambda_l$ 
is the laser beam polarisation) in the high-$x$ region. By scanning
over the different choices of the laser beam and initial electron
polarisation, we find that the best sensitivity is realised for the cases 
$(+,-,-,-)$ and $(-,+,-,-)$, where these represent the polarisations 
$(\lambda_{e1}, \lambda_{e2}, \lambda_{\ell 1}, \lambda_{\ell 2})$. 
In Table 2, we have displayed
the limits on $M_S$ obtained with the first of these choices for
the initial electron and laser beam polarisations. With this choice
of polarisations, we find that the limits can improve by about 20\%
as compared to the unpolarised case.

\begin{figure}[ht]
\vspace*{4.0in}
\includegraphics{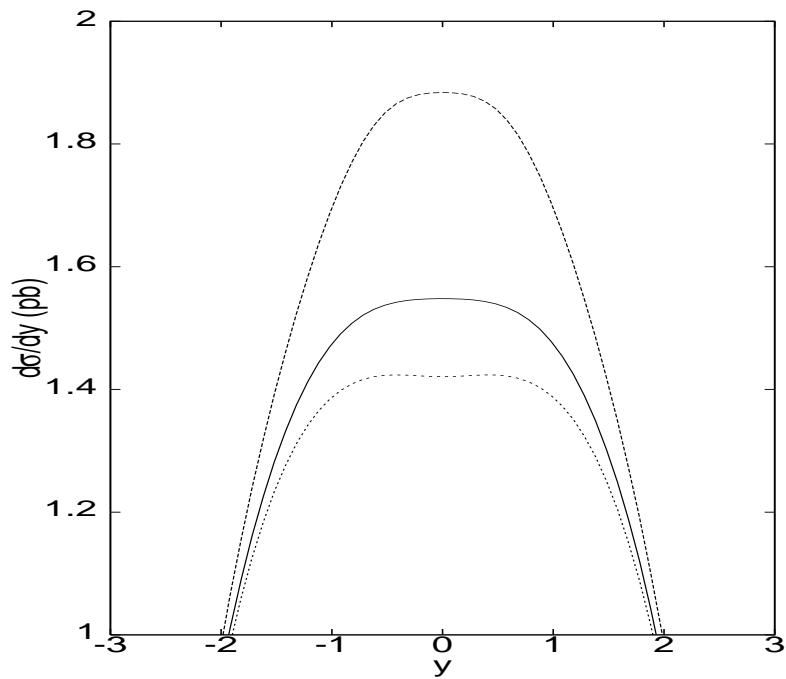}
\caption{\footnotesize\it The $y$ distribution for $\sqrt{s}_{ee}=1$
TeV for the unpolarised case. The solid line is the SM cross-section,
while the lines above and below the SM curve are for the SM+NSM cross-section
with $M_S=1$ TeV and $\lambda=-1$ and $\lambda=1$, respectively.} 
\end{figure}

In Fig.~1 we have plotted the rapidity distribution for $\sqrt{s}_{ee}=1$ 
TeV, in order to consider whether the use of differential quantities
will further enhance the sensitivity of the process under consideration
to the effects of the new physics. We find that the difference between
the SM and the SM$+$NSM distributions to be quite significant, especially
when we concentrate in the central regions of rapidity. Though we refrain
from making a quantitative estimate of the bound that would result (for
such an estimate would be premature without knowing further experimental 
details), it is clear from Fig.~1 that the rapidity distribution can
be used to improve the bound that would result from looking only at
the integrated cross-section.

We have shown that virtual effects of spin-2 Kaluza Klein exchange can
significantly affect dijet production in $\gamma \gamma$ collisions at
the NLC and so this process can be used to probe the physics of large 
extra dimensions. The range of $M_S$ values that can be probed with this 
process is of the order of 2.5 -- 6.5 TeV using the information on 
integrated unpolarised cross-sections. For the case of polarised
photon beams, the sensitivity is enhanced by about 20\%. Using
differential informations, like rapidity distributions may help
to further enhance the sensitivity of this process.

\end{document}